\documentclass[preprint,epsfig,floats,aps]{revtex4}
\usepackage{epsfig}
\begin{document}
\topmargin=-0.3cm
\title{Constrained Molecular Dynamics simulation of the Quark-Gluon Plasma}

\author{S.~Terranova and A. Bonasera 
 \footnote{Email: terranova@lns.infn.it; bonasera@lns.infn.it}}
\affiliation{
  Laboratorio Nazionale del Sud, Istituto Nazionale Di Fisica Nucleare,
      Via S. Sofia 44, I-95123 Catania, Italy \\
}

\begin{abstract}                

We calculate the Equation of State of a quark system interacting through a 
phenomenological potential: the Richardson's potential, at finite baryon density and zero 
temperature. In particular we study three different cases with different quark
masses($u$ and $d$),
and different assumptions for the potential at large distances.
We solve molecular dynamics with a constraint due to Pauli blocking and find  
evidencies of a phase transition from "nuclear" to "quark matter", which is analyzed also 
through the behaviour of the $J/\Psi$ embedded in the quark system. We show that the $J/\Psi$ particle behaves as an order parameter.
\end{abstract}

\maketitle


\section{Introduction}
The production of a new state of matter, the Quark-Gluon Plasma (QGP), is one of the open
problems of modern physics.
Theoretically Quantum Chromodynamics (QCD) predicts such a state, QGP, but it can be 
applied only to some limited
cases such as quark matter at zero density and high temperatures.
Experimentally such a system can be obtained through ultra-relativistic heavy ion 
collision (RHIC) at CERN and at Brookhaven \cite{wong}.
QGP can be formed in the first stages of the collisions, and can be studied 
through produced secondary particles.

Some features 
of the quark matter can be revealed by studying the properties of hadrons in a dense medium.
The particle $J/\Psi$ is a good candidate because the formation of the QGP might lead
to its suppression\cite{Jpsi}.
 
In this work we propose a semiclassical model which has an EOS resembling the well known 
properties of nuclear matter and predictions to the QGP at zero temperature and finite baryon 
densities.
We simulate the nuclear matter composed of 
nucleons (which are by themselves  composite three-quark objects) and its 
dissolving into quark matter. In addition, for our system of colored quarks, we will 
show how the color screening
is related to the lifetime of a particle $J/\Psi$ in the medium.
In particular we will see that the lifetime of the $J/\Psi$ as function of density behaves
as an order parameter.   Having a model which simulates the QGP might be useful when
dealing with finite and, possibly, out of equilibrium systems.  Infact dynamics and finite
size effects might wash completely or hide a phase transition.  The goal of our microscopic
simulations is to help find unambiguos signals of the occurrence of the phase transition.  In this
work we will show that indeed using some phenomenological potential and with suitable chosen quark masses we can obtain an EOS which has some features 
 of nuclear matter and its transition to the QGP.  Infact we stress that two systems having a similar EOS will behave the same.  An important ingredient of our approach is a constraint to satisfy the
Pauli principle.  The approach dubbed Constraint Molecular Dynamis (CoMD) as been 
successfully applied to relativistic and non relativistic \cite{bon2000,papa} heavy ion collision and plasma physics as well \cite{fus03}.

The paper is organized as follows: in Sect.II we introduce the method, molecular dynamics 
with a constraint for fermions, CoMD.
In Sect.III we apply the method to calculate the  Equation of State, with an arbitrary cut-off 
in the potential.
In Sect.IV we  use a screened linear potential and we calculate the EOS.
In Sect.V there is a brief summary. 

\section{Numerical Method}
We use  molecular dynamics with a constraint for a Fermi
system of quarks with colors.
The color degrees of freedom of quarks are taken into account through the Gell-Mann 
matrices and their dynamics is solved classically, in  phase space, following
the evolution of the distribution function.  
Starting from quarks degrees of freedom, some dynamical approaches 
have been proposed in \cite{bon99,maru00,mosel} based on the 
Vlasov equation \cite{repo,land1}, and/or molecular dynamics type approach. 
Of course, in such approaches it is important to get quark 
clusterization and the correct properties of nuclear matter (NM) at 
the ground state (gs) baryon density $\rho_0\sim0.15\ {\rm fm}^{-3}$\cite{pov}.  
However the property of ground state nuclear matter, 
together with the high density phenomena, is not sufficiently 
studied from the point of view of quarks degrees of freedom. 

In our work, the quarks interact through the Richardson's potential $V({\bf r}_i,{\bf r}_j)$: 
 \begin{eqnarray}
 V({\bf r}_{i,j})=3\sum_{a=1}^{8}\frac{\lambda_i^a}{2}\frac{\lambda_j^a}{2}\left[
\frac{8\pi}{33-2n_f}\Lambda(\Lambda r_{ij}-\frac{f(\Lambda r_{ij})}
{\Lambda r_{ij}})\right],
 \end{eqnarray}
and\cite{rich}
 \begin{eqnarray}
f(t)=1-4 \int{\frac{dq}{q}\frac{e^{-qt}}{[{\rm ln}(q^2-1)]^2+\pi^2}} .
 \end{eqnarray}
$\lambda^a$ are the Gell-Mann matrices. We fix the number of flavors 
$n_f=2$ and
the parameter $\Lambda=0.25$ GeV,($\hbar,c=1$) onless otherwise stated. 
Here we assume the potential to be dependent on the relative coordinates only.
The first term is the linear term, responsible of the confinement, the second term is 
the Coulomb term \cite{bon299}.

The exact (classical) one-body distribution function $f({\bf r},{\bf p},t)$ 
satisfies the equation \cite{land1}: 
   \begin{equation}
\partial_{t} f+\frac{\bf p}{E}\cdot 
\overrightarrow{\nabla} _{\bf r} f
-\overrightarrow{\nabla}
_{\bf r} U\cdot \overrightarrow{\nabla} _{\bf p} f=0 ,
\label{lv}
\end{equation}
where $E=\sqrt{p^2+m_q^2}$ is the energy,  
$m_q$ is the $(u,d)$ quark mass and $ U=U({\bf r})=\sum_{j} V({\bf r},{\bf r}_j)$. 
Numerically the  equation (3) is solved by writing the one body distribution
function for each particle $i$ through the delta function :
\begin{equation}
\label{efer}
 f_{i}({\bf r},{\bf p},t) =\sum_{\alpha=1}^{Q}\delta({\bf r}-{\bf r}_\alpha) \delta({\bf p}-{\bf p}_\alpha) ,
\end{equation}
where $Q=q+\bar q$ is the total number of quarks ($q$)  and antiquarks ($\bar q$)  (in this 
work $\bar q=0$). 
 
Inserting this expression in the exact equation (3) gives the Hamilton's equations:
\begin{equation}
\label{efer} 
\frac{d{\bf r}_i}{dt}=\frac{{\bf p}_i}{E_i} ,
\end{equation}
\begin{equation}
\label{efer}
\frac{d{\bf p}_i}{dt}=-\overrightarrow{\nabla}_{{\bf r}_{i}} U({\bf r}).
\end{equation}
Hence we must solve these equations of motion for our system of quarks.

Initially we distribute randomly the quarks in a box of side $L$ in coordinate space and
in a sphere of radius $p_f$ in momentum space.   
$p_f$ is the Fermi momentum
estimated in a simple Fermi gas model by imposing that a cell in
phase space of size $h=2 \pi$ can accommodate at most $g_q$ identical quarks
of different spins, flavors and colors.
$g_q=n_f\times n_c\times n_s$ is the degeneracy number, $n_c$ is the number
of colors (three different colors are used: red, green and blue ) hence $n_c=3$;
$n_s=2$ is the number of spins\cite{wong}. 
 
A simple estimate gives the following relation between the density of quarks with colors,
$\rho_{qc}$, and the Fermi momentum:
\begin{eqnarray}
\rho_{qc}=\frac{3n_s}{6\pi^2}p_f^3 
 \end{eqnarray}

We generate many events and take the average over all events in each cell on the phase space.
For each particle we calculate the occupation average, i.e. the probability that a cell in the
phase space is occupied.

To describe the Fermionic nature of the system we impose that average occupation for each particle is less or equal to 1 $(\bar{f_i}\leq 1)$.

At each time step we control the value of average distribution function and consequently we 
change the momenta of particles by multiplying them for a quantity $\xi$:
$P_i=P_i\times \xi$.
$\xi$ is greater or less than 1 if $\bar{f_i}$ is greater or less than 1 respectively; 
which is the $constraint$ \cite{papa}.  

With this procedure the basic quantities describing the system like: total energy, average occupation and order parameters(they will be described in this section below) after 
a given time will reach  stationary values.
We can see this in Fig.~1 in a typical case with $\rho_B =5.995 {\rm fm}^{-3}$.
We have repeated the calculations, in the same conditions, but two different starting points,
i.e. from quarks with colors randomly distributed, i.e. QGP (left) and from quarks condensed 
in clusters of three with different colors, i.e. nucleons (right), respectively.
We can see that in both conditions, the system will reach the same saturation values, though at
different times. We stress that this behaviour is independent of the density.


\begin{figure}[ht]
\centerline{\hspace{-0.5in}
\epsfig{file=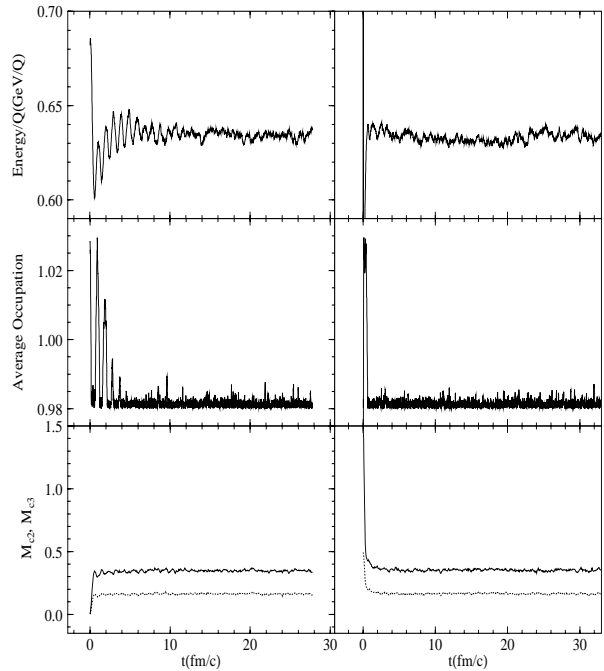,width=3.5in,height=4.0in,angle=0}}
\vspace{0.1in}
\caption{Time evolution of energy for quark(top panel), average occupation(middle panel) and 
reduced order parameters (bottom panel), with different starting point: QGP(left panels) and
nucleon(right panels).} 
\label{pau_pi}
\end{figure}
 
In the middle panel we display the time dependence of the average occupation, 
it is greater than 1 when we initially distribute 
randomly the quarks in the box 
and later it becomes nearly  to 1 at saturation.

We define an order parameter to check the order of a phase transition
if any. 
It is defined through the Gell-Mann matrices as\cite{bon2000}:
\begin{eqnarray}
 M_{c3}=\frac{1}{N}\sum_{i=1}^{N}\sum_{a=3,8}\lambda_j^a \lambda_k^a+ \lambda_i^a\lambda_j^a
 +\lambda_i^a \lambda_k^a= M_{c2}+\frac{1}{N}\sum_{a=3,8}\lambda_j^a \lambda_k^a
 + \lambda_i^a\lambda_k^a ,
\end{eqnarray}
where $j(i)$ and $k(i)$ are the two quarks closest to the quark $i$.
$M_{c2}$ is the reduced order parameter which gives the color of the particle $j$ 
closest to a particular quark $i$.

In Fig.~1 (bottom) we show the time evolution of $M_{c2}$ and $M_{c3}$ when the quarks are 
initially randomly distributed in the system (left) and when they
 are clusterized in nucleons (right),
at the same conditions as above.
The  saturation values are equal in both cases, but the initial ones are different, near to $0$ for the first case and near to $3/2$ for the second.
These values are typical for QGP and a system of nucleons respectively, as we will show
 later in this section.
   
To better understand the clusterization of colored quarks we also define an higher order parameter 
${M}_{c4}$ related to the colors of the $4$ closest quarks:
\begin{eqnarray}
 M_{c4}=\frac{1}{N}\sum_{i=1}^{N}\sum_{a=3,8}\lambda_j^a \lambda_k^a+ \lambda_i^a\lambda_j^a
 + \lambda_j^a\lambda_l^a+\lambda_i^a \lambda_k^a+\lambda_k^a \lambda_l^a
 + \lambda_i^a\lambda_l^a ,
\end{eqnarray}
where $l(i)$ is the third quark closest to the particular quark $i$.

We normalize the order parameters in this way:
\begin{eqnarray}
\widetilde{M}_{c2} &=& \frac{2}{3}[M_{c2}+1],\\
\widetilde{M}_{c3} &=& \frac{2}{9}[M_{c3}+3], \\
\widetilde{M}_{c4} &=& \frac{2}{15}[M_{c4}+6].
\end{eqnarray}

From the properties of the 
Gell-Mann matrices \cite{morpurgo} it is easy to derive the following results for the order parameters: 
if the three closest quarks have different colors then $\widetilde{M}_{c2}=1$ ($M_{c2}=1/2$),
$\widetilde{M}_{c3}=1$ ($M_{c2}=3/2$) and $\widetilde{M}_{c4} = 1$ ($M_{c2}=3/2$), in fact the fourth quark will have the same color
of one of the first three, in this case we have isolated white nucleons. 
This case is recovered in the calculation at small densities, 
where the system is locally invariant for rotation in color space.

If four closest quarks have the same color 
$\widetilde{M}_{c2} = \widetilde{M}_{c3} = \widetilde{M}_{c4}= 0 $ 
we have a condition that we call EXOTIC COLOR CLUSTERING. Also in this case the system is locally invariant for rotation in color space.
The corresponding potential energy is very large and repulsive. 

If the three closest quarks have two different colors, independently of the color of the two closest quarks,
i.e. the color of closest particle to quark $i$ is randomly chosen,
we have the Quark Gluon Plasma, hence: 
 $\widetilde{M}_{c2} = \widetilde{M}_{c3} =\frac{2}{3}$.
In this state $\widetilde{M}_{c4}$ can assume three different values:
$\frac{4}{5} $; $ 1 $; $\frac{3}{5} $ ,
according to the colors of the four closest quarks and number of pairs of different color.
If we have two pair of quarks with the same color (ex: $rggr$) $\widetilde{M}_{c4} =\frac{4}{5} $,
if the quarks have three 
different colors and two of the first three have two different colors (ex: $rggb$) $\widetilde{M}_{c4} =1$,
instead if three of the four closest quarks have the same color but the first three have two 
different colors (ex: $rggg$) $\widetilde{M}_{c4} =\frac{3}{5} $.
In the next sections we will better analyze  these quantities in different conditions.
We note that in the QGP case the system is globally invariant for rotation in color space.

To test for a signature of the various states of matter we studied the behaviour of  a pair of quarks $c$ and 
$\bar{c} $ embedded in the system, with $m_{c}=m_{\bar{c}}=1.37$ GeV.
For  each density we calculate the lifetime of the $J/\Psi$ particle, through her survival
probability in the system.  The $J/\Psi$  embedded in matter might split essentially for
two reasons. The first is that the internal kinetic energy of the $c$, $\bar{c} $ is large compared
to their mutual attraction (this is true in the case where the interaction is neglected-which we
will discuss below). The second, most important reason is that other quarks interact with 
the initial bound pair eventually splitting it.  Intuitively, it is clear that the splitting occurs
faster at higher densities where the $c$ ,$\bar{c} $ pair interacts with many other lighter quarks.
The survival probability is related to the total number of pairs $c$ and 
$\bar{c} $ that stay bound after they are inserted in our saturated system of $u$ and $d$ quarks.We have fitted the $J/\psi$ survival probability with the expression:
\begin{equation}
  P_{sur}(t)=exp[-(t-t_D)/\tau]
\end{equation}
and
\begin{equation}
  t_{sur}=t_D+\tau
\end{equation}
where $t_D$ is the delay time of the $J/\psi$ before the probability  
exponentially decreases, and $t_{sur}$ is the lifetime of $J/\psi$ in the system, similar to fission 
\cite{toshiki}.
A typical example of the fit is given in Fig.~2, where the dotted line is an example of the real distribution and the full line is obtained through Eq.(14).


\begin{figure}[ht]
\centerline{\hspace{-0.5in}
\epsfig{file=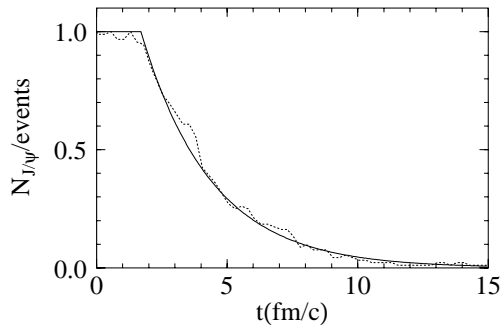,width=3.in,height=2.in,angle=0}}
\vspace{0.2in}
\caption{Time evolution of surviving $J/\Psi$s.} 
\label{pau_pi}
\end{figure}

In our study of the equation of state of quark matter at various baryon densities 
we look for some evidencies of phase transition to QGP 
also through the properties of the meson $J/\Psi$ in the medium.  This is important because we
want to see if the $J/\Psi$ can tell us about the occurrence and the order of the phase transitions. 
In future works for finite systems we want to test if the properties of the $J/\Psi$ remain.  Infact,
in an infinite systems there is all the time for dissolving the $J/\Psi$, but in a rapidly expanding
 QGP this might not occur also because of the relatively large charm masses. 

\section{Results with cut-off}
When quarks, objects of different colors, are embedded in a dense medium such as in nuclear
matter, the potential becomes screened in a similar fashion as ions and electrons in condensed 
matter. This is the $Debye$ $screening$ \cite{wong,land1}.

The  screening can be obtained through the use of a Debye radius screening in the interaction, 
as  will be discussed in the next section. In this section, the screening is produced directly, through the interaction of colored quarks.
But our system is not really an infinite system, like nuclear matter, 
and this screening is insufficient to screen the linear potential and avoid its 
divergence for $r\rightarrow\infty$,
hence we introduce a cut-off for the potential.
The cut-off is a free parameter, when quark distances are greater then the cut-off, the 
interaction is equal to zero.   Of course we are aware that by using a cut-off in the linear term
the confinement property of the quarks might be lost.  Neverthless we will show that this prescription leads to interesting effects which might be used in finite system studies.  Furthermore
the cut-off is relatively large thus it takes a considerable energy to have isolated quarks.  

In Fig.~3 we plot some quantities related to the case with small quark masses,
$m_u =5$ MeV, $m_d =10$ MeV   and a cut-off of $3$ fm.


\begin{figure}[ht]
\centerline{\hspace{-0.5in}
\epsfig{file=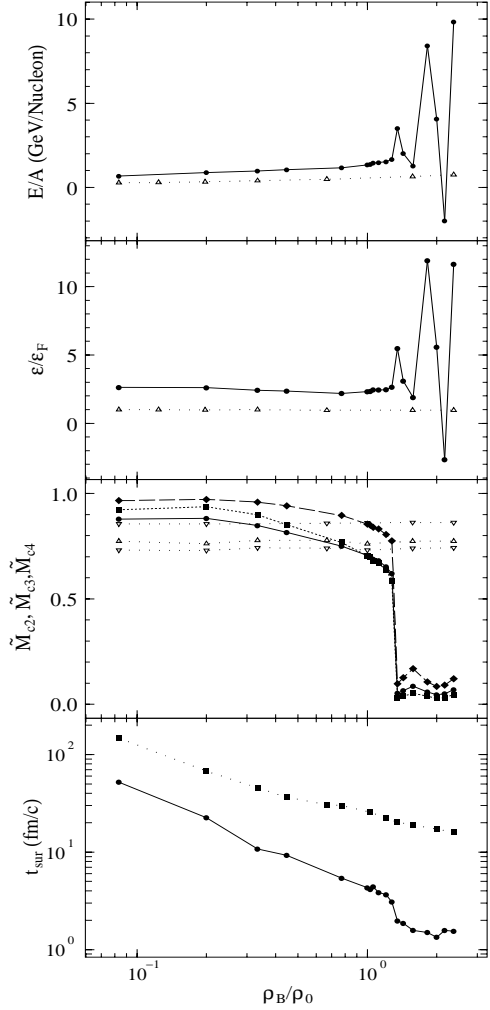,width=3.in,height=5.7in,angle=0}}
\vspace{0.2in}
\caption{Energy per nucleon (top panel),
energy density(2nd to top), normalized order parameters
(middle panel) and time survival of $J/\psi$ (bottom panel) versus density
divided by the normal density $\rho_0 $, for $m_u =5$ MeV, $m_d =10$ MeV 
and cut-off$=3fm$.} 
\label{pau_pi}
\end{figure}

The energy per nucleon and corresponding energy density in units of the $\varepsilon_F$ (energy
density for a Fermi gas \cite{wong}) (top panels) versus baryon density divided
by the normal density $\rho_0 $, have a very irregular behaviour that we 
can explain through the order parameters (third panel).

For small densities the quarks are condensed in clusters of three different colors,
the system is locally white (isolated white nucleon). 
The  normalized order parameters: $\widetilde{M}_{c2}$ (circles), $\widetilde{M}_{c3}$ (squares)
and $\widetilde{M}_{c4}$ (rombs) are near 
to $1$, hence the two closest particle to quark $i$ have different colors and consequently the third closest quark to 
$i$ has the same color of one of the first two or of $i$, 
$\widetilde{M}_{c2} = \widetilde{M}_{c3}=\widetilde{M}_{c4}\simeq 1$ . 
At  higher densities, the quarks are not in clusters but randomly distributed,
$\widetilde{M}_{c2} = \widetilde{M}_{c3}\simeq \frac{2}{3} $ and $\widetilde{M}_{c4}\simeq \frac{4}{5}$, 
and we have the QGP ($\rho_B/\rho_0\sim 1.2 $).
But the system does not stay in this state, it prefers the exotic color clustering state,
where at least the four closest quarks have the same color.
At density about $1.4\sim2.4$ times the normal density 
$\widetilde{M}_{c2} = \widetilde{M}_{c3} = \widetilde{M}_{c4} \simeq 0 $.
The system reaches this state through a first order phase transition \cite{wong} at about 1.3 times
the normal nuclear matter density. In the figure relative to the energy per nucleon, the transition is signaled by a discontinuity at the same density.
The other discontinuities at larger densities ($\rho_B/\rho_0 > 1.5 $), are probabily due to the clusterization of more than four quarks of the same color.  The reason for the phase transition at
such small density is due to the small quark masses and to the large cutoff radius.  As we will
show more in detail below we can change those values and change not only the density where
the transition occurs but also the order of the transition, if any.

In the present conditions, the linear term becomes
very large and positive, hence the attraction between different clusters of quarks with different color distant in  space prevails respect to repulsion between charges of the same color in each cluster, this explains the large values of the energy and consequently of energy density. For instance,  we might have the formation of  three red quarks cluster and these are attracted by an analogus three green quarks cluster.  

It is the linear term that produces this very irregular behaviour, in fact 
repeating  the same calculations with the Coulomb term only, triangles
in Fig.~3, we obtain a constant contribution, not only to the energy, but also to the order parameters. The Coulomb term produces a permanent clusterization among quarks  
which prevents them to reach the  ideal QGP state.
Infact the density dependence of  the Coulomb term is similar to the Fermi energy term.  The difference between the two terms   depends on the  $\alpha_s $ value, where $\alpha_s$ is the strong constant coupling defined in the potential through the $\Lambda $ parameter.  
When the linear term is included it prevails respect to the Coulomb one and the system stays in exotic color clustering
state, $\widetilde{M}_{c2}  = \widetilde{M}_{c3} = \widetilde{M}_{c4}= 0$.
 
In Fig.~3 (bottom) we plotted the lifetime of $J/\Psi$ versus baryon density divided by $\rho_0$ (full line).
When the density increases, the lifetime decreases because 
it is more probable that a particle of different flavour gets in between a $c \bar{c} $ pair 
and breaks the bond, then the number of survival $J/\Psi$ in the medium decreases fastly and we have small values of $t_{sur}$.

$t_{sur}$ behaves similarly to an order parameter, infact it 
has a jump just where we found  the phase transition ($\rho_B/\rho_0\sim 1.3$).  
Analyzing the particle $J/\Psi$ in the medium turning off the interaction (squares in Fig.~3 bottom), gives a different behaviour, i.e. a monotonic decrease with density. The survival time in the medium is always larger than that with interaction, because the
forces break more easily the bonds between particles($c$,$\bar{c}$ quarks).
After the jump we notice a saturation of the surviving probability, again similarly to the order parameter. We would like to note that even though there is not much similarity between the EOS obtained 
here and nuclear matter with its transition to the QGP, it was the first case we studied and its
 features are quite general as we will see in the following.  Infact it would suffice a simple scaling 
around the critical density to compare to the other systems.  We notice also that even if the
$J/\Psi$ is usually studied at zero density and finite temperatures we expect a similar behaviour
to the one discussed here, with the Fermi motion playing the role of the temperature.


\begin{figure}[ht]
\centerline{\hspace{-0.5in}
\epsfig{file=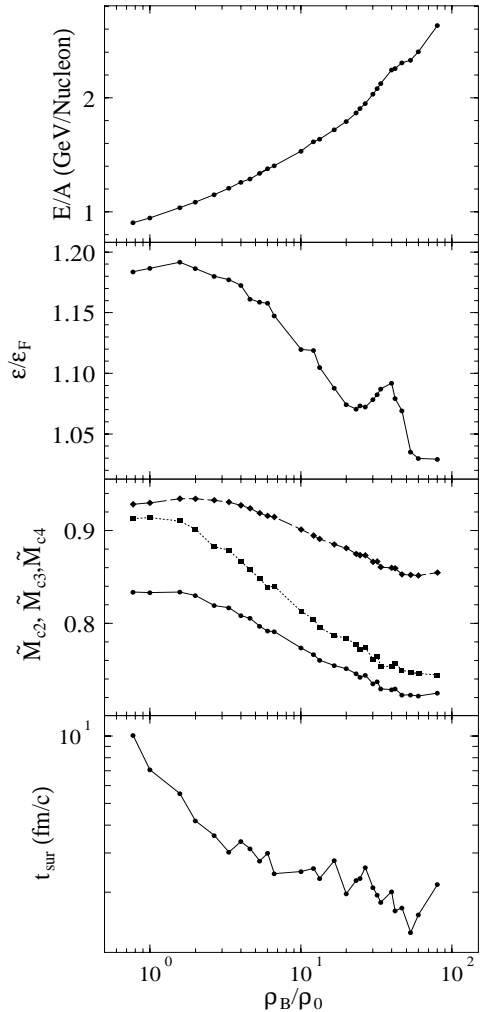,width=3.in,height=5.7in,angle=0}}
\vspace{0.2in}
\caption{Energy per nucleon (top panel),
energy density(2nd to top panel), normalized order parameters
(middle panel) and time survival of $J/\psi$ (bottom panel) versus density
divided by the normal density $\rho_0 $ for $m_u =m_d =180$ MeV 
and a cut-off of $1.26$ fm.} 
\label{pau_pi}
\end{figure}

In order to study the sensitivity of the results to the input parameters,
we have repeated the calculations with  
$m_u =180$ MeV, $m_d =180$ MeV and a cut-off equal $1.26$ fm.
The quark masses are chosen to reproduce the energy per nucleon  of nuclear matter 
at normal density \cite{pov}.  
In Fig.~4  where we plot the same quantities of Fig.~3 (symbols have the same meaning), we can see a 
behaviour more regular than previously.
At very high densities (almost 45 times the normal density), 
in the figure relative to the energy per nucleon(top) 
we see a flex, probably indicating a second order phase transition, which becomes a change of slope in energy density(figure below). 
The normalized order parameters are always positive, i.e.
it never happens that three equal quark color states are on average in the same region in
$r$ space. At low densities actually they never  
reach the value of $1$ (nucleons). Which implies that our potential is insufficient to get a good 
clusterzation, in fact we do not obtain the minimum in energy per nucleon indicating a condition 
of stability for the system.

Values of order parameters at high density are always larger than $2/3$ for $\widetilde{M}_{c2}$ and $\widetilde{M}_{c3}$, and $4/5$ (one of the possible values of $\widetilde{M}_{c4}$ to have QGP). 
This indicates a residual clusterization between quarks of different color, which we associate to the semiclassical counterpart of pairing. Infact a residual attractive force especially due to the Coulomb term, couples quarks of different color. 

Also in this case we studied the behaviour of the $J/\Psi$ in the medium and we obtained a
regular behaviour for the lifetime, Fig.4 (bottom), i.e. a fast decrease for
small densities, and after about 7$\rho_0$ a slow decrease with some fluctuation around $1\sim2fm/c$.
The lifetime of $J/\Psi$ behaves again similarly to the order parameters. 
In conclusion, we can see how changing quark masses and cut-offs we do not have the exotic color clustering 
and a first order phase transition, but a probable second order phase transition to QGP.  

\begin{figure}[ht]
\centerline{\hspace{-0.5in}
\epsfig{file=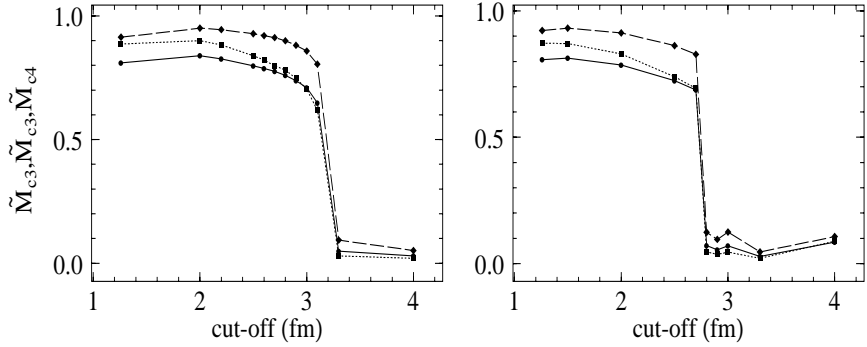,width=5.in,height=2.5in,angle=0}}
\vspace{0.2in}
\caption{Reduced order parameters vs cut-off for $m_u =5$ MeV, $m_d =10$   at two different densities.} 
\label{pau_pi}
\end{figure}

It is clear that the  cut-off value changes the transition  point  consequently.
 In Fig.~5 we plot the reduced order parameters $\widetilde{M}_{c2}, \widetilde{M}_{c3}, \widetilde{M}_{c4}$ versus cut-off at density $\rho_B=\rho_0$(left) and
$\rho_B=0.3558 fm^{-3}$ ($\sim 2.3 \rho_0$) (right) for $m_u =5$ MeV, $m_d =10$ MeV.
Changing the cut-off values from $1.26$ to $4 fm$ gives a phase transition from QGP to exotic color clustering state, at $3.1 fm$  for smaller density and at $2.7 fm$  for $\rho_B=0.3558 fm^{-3}$, while the system was in a nucleonic state for small cut-off values for both cases.


\begin{figure}[ht]
\centerline{\hspace{-0.5in}
\epsfig{file=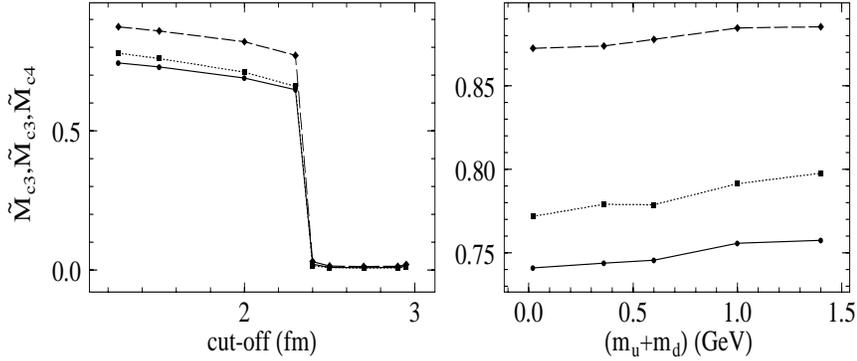,width=5.in,height=2.5in,angle=0}}
\vspace{0.2in}
\caption{Reduced order parameters vs cut-off for $m_u =m_d =0.18 MeV$ and vs quark masses for cut-off$=1.26fm$.} 
\label{pau_pi}
\end{figure}

At $\rho_B=3.49 fm^{-3}$ ($\sim 23.26 \rho_0$) we calculated the order parameters versus cut-off, Fig.~6, for $m_u =m_d =180$ MeV(left) and versus quark masses for a cut-off$=1.26fm$(right).

As the previous case with smaller quark masses Fig.~5, for small cut-off values we have typical values of a nucleonic state. When the cut-off increases we have a phase transition from QGP to exotic color clustering state, at a cut-off $=2.3 fm$. 
If we analize the reduced order parameters versus quark masses we find an almost constant
behaviour .   Very large variations of quark masses correspond little variations of the $\widetilde{M}_{c2}, \widetilde{M}_{c3}, \widetilde{M}_{c4}$ values, 
 hence it is the cut-off mainly responsible of the phase transition at different densities. 

This suggests  to use a different cut-off value to change the point of transition. In  Fig.~7 we plot the energy per nucleon, energy density, normalized order parameters and time survival of $J/\psi$,
for $m_u =5$ MeV, $m_d =10$ MeV and cut-off$=2fm$.

\begin{figure}[ht]
\centerline{\hspace{-0.5in}
\epsfig{file=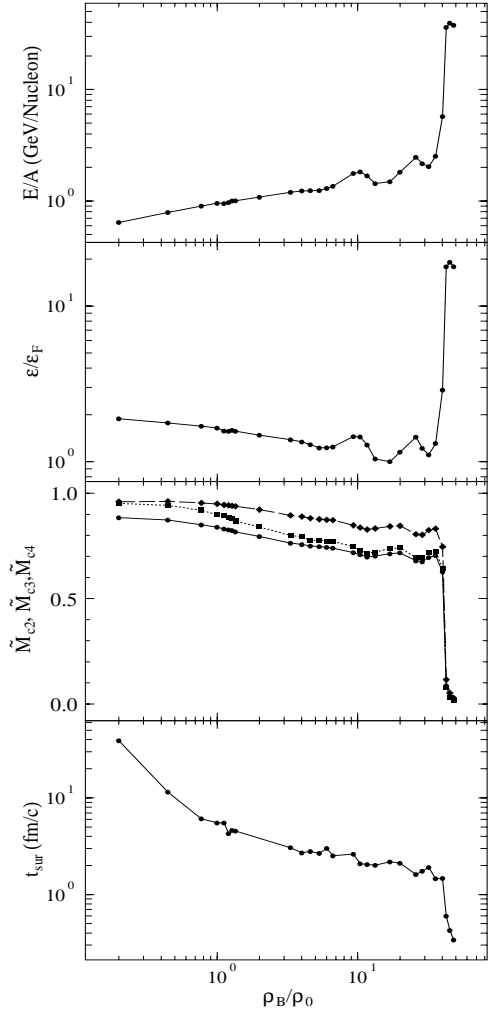,width=3.in,height=5.7in,angle=0}}
\vspace{0.2in}
\caption{Energy per nucleon (top panel), energy density(2nd to top panel), normalized order parameters
(middle panel) and time survival of $J/\psi$ (bottom panel) versus density
divided by the normal density $\rho_0 $ for $m_u =5$ MeV,$m_d =10$ MeV and cut-off$=2fm$.} 
\label{pau_pi}
\end{figure}

At $\rho_B=\rho_0$ the total energy per nucleon(top) has a value similar to the typical value of nuclear matter but we do not have the minimum. Important physics is infact missing in our approach   to describe the system of quark at small densities, i.e. the nuclear part.
At $\rho_B=10\rho_0$ the energy per nucleon and energy density(2nd to top) display
 some fluctuation  near to the QGP state.  In fact the value of the reduced order parameters (middle figure) are close to $2/3$ for $\widetilde{M}_{c2}$, $\widetilde{M}_{c3}$ and  $4/5$ for $\widetilde{M}_{c4}$.
Increasing the density further  gives a first order phase transition to exotic color clustering state at $\rho_B\sim 40\rho_0$, as signaled  by the reduced order parameters $\widetilde{M}_{c2}=\widetilde{M}_{c3}=\widetilde{M}_{c4}\simeq 0$ and the  large increase of energy and energy density.
The $J/\Psi$ particle displays  some fluctuation around $10\rho_0$ and a jump around $40\rho_0$, similar to the reduced order parameters.

This equation of state could be our initial condition to simulate a finite system and a collision between nuclei.

\section{Debye screening}
In this section, to have a good screening 
of the linear interaction we use a particular expression of 
the linear term obtained through the resolution of the Poisson
equation in one dimension \cite{land2}:
\begin{equation} 
\nabla^{2}\phi_{Lin}=-\sum_iq_i\rho_{qi}   
\end{equation}
   
where $\rho_{qi}$ is the linear density obtained in the  Thomas-Fermi approximation:
\begin{equation}
\rho_{q_i}\approx\frac{E_{F_0}}{{(6\pi^2})^{1/3}}
\left[\left(1-\frac{\phi_{Lin}}{E_{F_0}}\right)^{2}
-\frac{{m_q}^2}{E_{F_0}^2}\right]^{1/2}
\end{equation}

$E_{F_0}$ is the Fermi energy calculated at
large distances, where the field $\phi_{Lin}(r)\rightarrow 0$ and
$P_{F_0}=\left(\frac{6\pi^2}{g_q}\rho_{q0}\right)^{1/3}$,
i.e. we require that all the r-dependence
is contained in the field and the density 
reduces to the free one for large distances from a given 
quark.
Hence:
 $$
  \phi_{lin}=
  \left \{
 \begin{array}{lc} 
 \frac{K}{\chi_l}\exp(-\chi_l\:r_{ij}) &  \;\rho_q \neq 0  \\ 
  Kr_{ij} & \;\rho_q = 0
 \end{array}
 \right.
 $$

with
 \begin{equation}
 \chi_l^{2}=\frac{4}{\sqrt{3}}K(\frac{g_q}{6\pi^{2}})^{\frac{1}{3}}\frac{\sqrt{{P_F}^{2}
 +{m_q}^{2}}}{P_F} 
 \end{equation}

$\chi_l$ is the linear Debye inverse radius \cite{land2}, that for large densities goes to a constant.
Also the linear potential goes to a constant and not to zero like for a total screening,
but for distance larger than the Debye radius the potential is screened. 
 $K$ is the string tension defined through the 
$\Lambda$ parameter \cite{morpurgo}.  We stress that the confinement property is recovered in
 the zero density limit.


\begin{figure}[ht]
\centerline{\hspace{-0.5in}
\epsfig{file=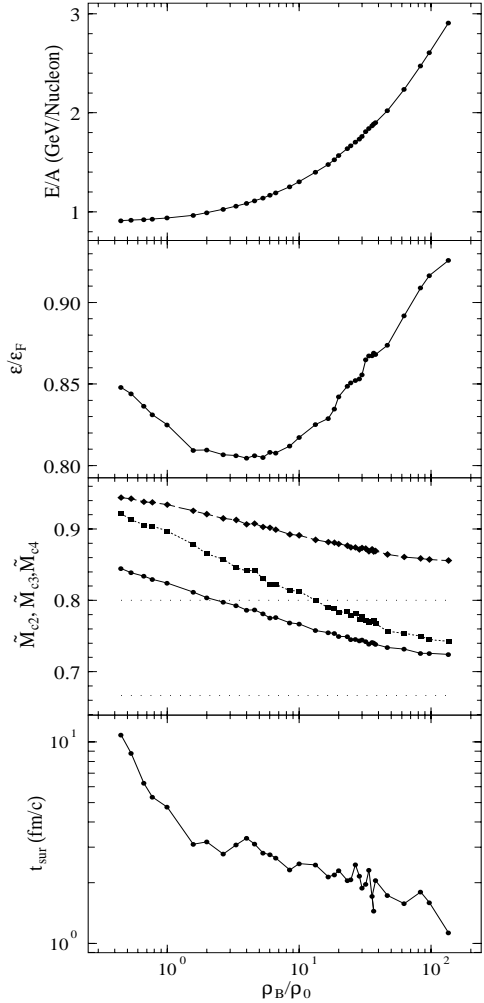,width=3.in,height=5.7in,angle=0}}
\vspace{0.2in}
\caption{Energy per nucleon (top panel),
energy density(2nd to top panel), normalized order parameters
(middle panel) and time survival of $J/\psi$ (bottom panel) versus density
divided by the normal density $\rho_0 $ for $m_u =m_d =324$ MeV.} 
\label{pau_pi}
\end{figure}

In Fig.~8 we plot some results with this potential and $m_u =m_d =324$ MeV. Also in this case
we choose the masses of quarks
to reproduce the energy per nucleon  of nuclear matter at normal density.

In the top figure
relative to energy per nucleon we found a minimum for low densities.
The energy density, for large densities, shows some fluctuations, which can be due to numerical fluctuations. 

The order parameters  are smooth functions of the density and also in this case they
 never  reach the  QGP values (dotted lines):
$\widetilde{M}_{c2} =\widetilde{M}_{c3} = \frac{2}{3} $ and $\widetilde{M}_{c4}= \frac{4}{5} $ (or other possible values), i.e. a residual clusterization remains. 

Probably it is an effect of the instability of quark pairs of different
colors that produces the instability at large energy density.
Also in this case the 
reduced order parameters are always positive and we never have  exotic color clustering.

The calculated lifetime of $J/\Psi$ in the medium 
versus density is shown in Fig.~8 (bottom) and it  behaves  like  the order parameters. 

\section{Summary}
In conclusion, in this work we have discussed a semiclassical molecular dynamics approach to 
infinite matter at finite baryon densities and zero temperature starting from a phenomenological 
potential that describes the interaction between quarks with color.  Pauli blocking, necessary for Fermions at zero temperature, is enforced through a constraint to the average one body 
occupation function.
Color degrees of freedom for quarks are  responsible for Debye screening,  even though we have 
adopted some prescription mainly for numerical reasons to screen the linear term at large 
distances.
Depending on parameters for the quark masses and the potential, we obtain EOS which
exhibit a first, second or a simple cross over to the QGP.  We stress that these transitions are
due to changes in the system symmetries.  Infact we have a local invariance for rotation at low
densities i.e. for nucleons.  This means that we can rotate locally the color of the quarks
with no change in the energy.  The QGP displays a global invariance, i.e. we can change randomly the quarks colors anywhere in the system without changing the system properties. Exotic color clustering is also a local property, infact we can change the color randomly but only within a cluster of identical colors.  It is the breaking of these symmetries which gives the phase transitions. A suitable physical observable for the phase transition could be the $J/\Psi$, infact we have shown that for infinite systems it behaves like an order parameter and it is also able to distinguish between  a first order and a second order (or a cross-over) phase transition.  Finite size and dynamical studies within the model proposed will revail if such a property remains.  Those
studies could also give indications on the possibility that the phase transitions are washed by 
finite sizes effects.  Also other indicators of a phase transitions such as intermittency can be easily
studied in the framework of our model.

\end{document}